\documentclass[12pt,letterpaper]{article}
\usepackage{amssymb}
\usepackage{amsmath}
\usepackage{endnotes}
\usepackage[onehalfspacing]{setspace}

\input{tcilatex}
\let\footnote=\endnote

\begin{document}

\title{Errors-in-variables models: a generalized functions approach}
\date{Revised version of September 29, 2009 }
\author{Victoria Zinde-Walsh\thanks{%
The support of the Social Sciences and Humanities Research Council of Canada
(SSHRC), the \textit{Fonds\ qu\'{e}becois de la recherche sur la soci\'{e}t%
\'{e} et la culture} (FRQSC) is gratefully acknowledged. \ The author thanks
participants in the Cowles Foundation conference, the UK ESG, CEA 2009,
Stats in the Chateau and CESG meetings and P.C.B.Phillips, X. Chen, D.
Nekipelov, L.Wang and T.Wootersen for illuminating discussions and
suggestions. An anonymous referee provided a very thorough report containing
many important points and suggestions. The remaining errors are all mine.} \\
McGill University and CIREQ}
\maketitle

\begin{abstract}
Identification in errors-in-variables regression models was recently
extended to wide models classes by S. Schennach (Econometrica, 2007) (S) via
use of generalized functions. In this paper the problems of non- and semi-
parametric identification in such models are re-examined. Nonparametric
identification holds under weaker assumptions than in (S); the proof here
does not rely on decomposition of generalized functions into ordinary and
singular parts, which may not hold. Conditions for continuity of the
identification mapping are provided and a consistent nonparametric plug-in
estimator for regression functions in the $L_{1}$ space constructed.
Semiparametric identification via a finite set of moments is shown to hold
for classes of functions that are explicitly characterized; unlike (S)
existence of a moment generating function for the measurement error is not
required.

\end{abstract}

\bigskip

\bigskip

\begin{center}
\bigskip \pagebreak
\end{center}

\section{Introduction}

The familiar errors in variables model with an unknown regression function, $%
g,$ and measurement error in the scalar variable has the form 
\begin{eqnarray*}
Y &=&g(X^{\ast })+\Delta Y; \\
X &=&X^{\ast }+\Delta X,
\end{eqnarray*}%
where variables $X$ and $Y$ are observable; $X^{\ast }$ and $\Delta X,\Delta
Y$ are not observable. A widely used approach makes use of instrumental
variables.\ Suppose that instruments are available and $Z$ represents an
identified projection of $X$ on the instruments so that additionally $%
X^{\ast }=Z-U;$\ assume that\ $U$ is independent of $Z\ $(Berkson-type error
from using the instruments) and\ that\ 

\begin{eqnarray*}
\text{ }E[\Delta Y\left\vert Z,U\right. ] &=&0; \\
E[\Delta X\left\vert Z,U,\Delta Y\right. ] &=&0; \\
E(U) &=&0.
\end{eqnarray*}%
These assumptions\ were made by e.g. Hausman et al. (1991) who examined
polynomial regression. Newey (2001) added another moment condition for
estimation \ in semiparametric regression leading to two equations for
unknown $g$ and $F,$ the measurement error distribution (all integrals over $%
(-\infty ,\infty )):$ 
\begin{eqnarray*}
E(Y|Z &=&z)=\int g(z-u)dF(u); \\
E(YX|Z &=&z)=\int (z-u)g(z-u)dF(u).
\end{eqnarray*}%
Define $W_{y}(z)\equiv E(Y|Z=z);W_{xy}(z)\equiv E(Y(Z-X)|Z=z).$ The model
assumptions then can be considered in terms of classes of functions $%
W_{y},W_{xy},g$ and distribution $F$ that satisfy the equations: 
\begin{eqnarray}
W_{y}(z) &=&\int g(z-u)dF(u);  \label{W1} \\
W_{xy}(z) &=&\int (z-u)g(z-u)dF(u);  \notag
\end{eqnarray}%
we say that these functions $W_{y},W_{xy},g$ $,F$ satisfy model assumptions.
The functions $g$ and $F$ enter in convolutions; this motivates using
Fourier transforms (Ft). Fourier transforms:%
\begin{eqnarray}
\varepsilon _{y}(\zeta ) &=&Ft(W_{y}(\cdot ));  \label{Ft} \\
\varepsilon _{xy}(\zeta ) &=&Ft(W_{xy}(\cdot ));  \notag \\
\gamma (\zeta ) &=&Ft(g(\cdot ));  \notag
\end{eqnarray}%
the characteristic function is obtained as $\phi (\zeta )=\int e^{i\zeta
u}dF(u).$

Provided that for some subclass of functions Fourier transforms are well
defined, derivatives exist and the convolution theorem applies, (\ref{W1})
is equivalent to a system with two unknown functions, $\gamma ,\phi :$ 
\begin{eqnarray}
\varepsilon _{y}(\zeta ) &=&\gamma (\zeta )\phi (\zeta );  \label{Fy} \\
i\varepsilon _{xy}(\zeta ) &=&\overset{\cdot }{\gamma }(\zeta )\phi (\zeta ),
\label{Fxy}
\end{eqnarray}%
where $\overset{\cdot }{\gamma }=\frac{d\gamma }{d\zeta }.$ S. Schennach
(2007) (S) suggested that these equations can be justified for a wide class
of functions if one uses generalized functions, specifically, those in the
space of tempered distributions, $T^{\prime }$ (defined below in section 2.1)%
$^{1}$\footnotetext[1]{%
A referee pointed out that the usual notation for the space of tempered
distributions is $S^{\prime },$ but here we follow the notation in (S).}.

\textbf{Assumption 1. }The functions $g,W_{y},W_{xy}$ that satisfy the model
assumptions are such that each represents an element in the space of
tempered distributions, $T^{\prime }.$

Some examples of such functions are the class considered in (S, Assumption
1): functions such that $\left\vert g(x^{\ast })\right\vert ,\left\vert
W_{y}(z)\right\vert ,\left\vert W_{xy}(z)\right\vert $ are defined and
bounded by polynomials on $R.$ However, the assumption here allows for very
wide classes of functions. This class may be difficult to characterize
explicitly; the Assumption 1' below provides an important subclass of
locally integrable functions  in $T^{\prime }$.

Consider functions $b(t)$ for $t\in R$ that satisfy%
\begin{equation}
\int (1+t^{2})^{-l}\left\vert b(t)\right\vert dt<\infty \text{ \textit{for
some }}l\geq 0.  \label{cond}
\end{equation}

\bigskip \textbf{Assumption 1'. }The functions $g,W_{y},W_{xy}$ that satisfy
the model assumptions are such that each satisfies (\ref{cond}).

The functions $g,W_{y},W_{xy}$ that satisfy (\ref{cond}) satisfy Assumption
1. Any function in the space $L_{1}$ of absolutely integrable functions
satisfies Assumption 1' here but not the Assumption 1 in (S) unless the
function is everywhere bounded. While the assumption in (S) extends to
polynomial regression functions or distribution functions for binary choice
models (where Ft do not exist in the ordinary sense), a regression function
that is unbounded at some points is not allowed. There are cases where such
properties may arise, e.g. for some hazard functions, for liquidity trap;
the more general assumption here accommodates such cases.

Fourier transform is a continuous invertible operator in $T^{\prime }$, all
tempered distributions are differentiable in $T^{\prime }$ (thus $\overset{%
\cdot }{\gamma }$ is defined). Fourier transform of an ordinary function of
the type considered here may no longer be an ordinary function (e.g. $%
Ft(const)=\delta ,$ the Dirac delta-function that cannot be represented as
an ordinary function), and thus is not defined point-wise; thus the notation 
$\gamma (\zeta ),$ etc. for the $Ft$ in e.g. (\ref{Fy},\ref{Fxy}) which we
keep here for convenience refers just to the generalized function $\gamma $
without necessarily giving meaning to values at a point.

In the class of functions that satisfies the model assumptions denote by $A$
the subclass of functions $\left( g,F\right) $, by $A^{\ast }$ the subclass
of functions $(g);$ the mapping $P:A\rightarrow A^{\ast }$ is given by $%
P(g,F)=g.$ Denote by $B$ the class of functions $(W_{y},W_{xy})$. Equations (%
\ref{W1}) of model assumptions map $A$ into $B$ (mapping $M:A\rightarrow B);$
Fourier transforms map $B$ into $Ft(B),$ the class of functions that are
Fourier transforms of functions from $B;$ if equations (\ref{Fy},\ref{Fxy})
could be solved they would provide solutions $\phi ^{\ast }=$ $\phi I(\gamma
\neq 0)$ (where $I(A)=1$ if $A$ is true, zero otherwise) and $\gamma ^{\ast }
$ if $\phi \neq 0$; applying inverse Fourier transform would give $g^{\ast
}=Ft^{-1}(\gamma ^{\ast }).$ This sequence of mappings can be represented as
follows:

\begin{equation}
\underset{(g,F)}{A}\overset{M}{\rightarrow }\underset{(W_{y,}W_{xy})}{B}%
\overset{Ft}{\rightarrow }\underset{(\varepsilon _{y},\varepsilon _{xy})}{%
Ft(B)}\overset{S}{\rightarrow }\underset{(\phi ^{\ast },\gamma ^{\ast })}{Ft(%
\tilde{A})}\rightarrow \underset{(\gamma ^{\ast })}{Ft(A^{\ast })}\overset{%
Ft^{-1}}{\rightarrow }\underset{(g^{\ast })}{A^{\ast }}.  \label{chain}
\end{equation}%
If (\ref{chain}) provides the same result as $P$ so that $g^{\ast }\equiv g$
(and $\gamma ^{\ast }\equiv \gamma )$ then $g$ can be identified from the
functions $(W_{y},W_{xy})$ with the identification mapping 
\begin{equation}
M^{\ast }:B\rightarrow A^{\ast }  \label{M}
\end{equation}%
given by composition of the last five mappings in (\ref{chain}). The most
challenging part is in solving the equations to establish the mapping (for $%
\gamma ^{\ast }\equiv \gamma )$ 
\begin{equation}
S:\underset{(\varepsilon _{y},\varepsilon _{xy})}{Ft(B)}\rightarrow \underset%
{(\phi ^{\ast },\gamma )}{Ft(\tilde{A})}  \label{S}
\end{equation}%
Two additional assumptions are similar to those in (S) and are standard.

\textbf{Assumption 2}. The function $\phi (\zeta )$ is continuous,
continuously differentiable on $R$; and $\phi (\zeta )\neq 0.$

In terms of the model this implies a further condition that absolute moment
of $U$ exist.

\textbf{Assumption 3.} Support of generalized function $\gamma $ coincides
with $\left\vert \zeta \right\vert \leq $ $\bar{\zeta}$ where $\bar{\zeta}>0$
and could be infinite.

Under Assumptions 1-3 identification is possible as shown in Theorem 1 of
this paper; the theorem in (S) asserts an analytic formula (S, (13)) that
relies on a decomposition that may not hold.

When the errors-in-variables problem is examined in the space of tempered
distributions the corresponding (weak) topology is that of the space $%
T^{\prime };$ in that topology the mappings Ft, Ft$^{-1}$ are known to be
continuous, however, the mapping (\ref{S}) may be discontinuous, rendering
the identification mapping (\ref{M}) discontinuous as well thus implying
ill-posedness of the problem. One reason for this is that a too thin-tailed
characteristic function may mask high-frequency components in the Fourier
transform of the regression function. Theorem 1 here provides a condition
under which continuity obtains. When identification is provided by a
continuous mapping nonparametric plug-in estimation is possible as long as
Fourier transforms of the conditional moment functions can be consistently
estimated in $T^{\prime }$; this applies e.g. if the regression function is
in the $L_{1}$ space; Proposition 2 establishes this result.

Identification in classes of parametric functions requires that the mapping
from the parameter space to the function space be (at least locally)
invertible. (S) uses generalized functions to widen classes of parametric
functions for which identification is provided by a finite number of moment
conditions; in particular she expands classes of $L_{1}$ functions to which
the results of Wang and Hsiao (1995, 2009) apply and also allows sums of
such functions with polynomial functions, where before polynomial functions
were considered only by themselves in Hausman et al. (1991). Her results
rely on existence of a moment generating function for the measurement error
and use special weighting functions (some of which are improperly defined).
Here general classes of functions where such identification is achievable
are explicitly characterized rather than via existence of moments conditions
(as in S, Assumption 6), the requirement of a moment generating function\
for measurement error is avoided; appropriate weighting functions are given.

Section 2 deals with identification and well-posedness in the non-parametric
case. Section 3 examines identification for the semiparametric model. Proofs
are in Appendix A. Appendix B provides an explanation of the claims about
the main errors in (S).

\section{Non-parametric identification}

In the first part of this section known results on generalized functions
that confirm the existence and continuity of some of the mappings in (\ref%
{chain}) are provided, in particular, for the Fourier transform and its
inverse$.$ Other mappings, such as ($\ref{S})$ require special treatment
because they involve multiplication of generalized functions. Multiplication
in spaces of generalized functions cannot be defined (Schwartz's
impossibility result, 1954, see also Kaminski and Rudnicki (1991) for
examples) although there are cases when specific products are known to
exist. Here conditions under which some generalized functions can be
multiplied by some continuous functions to obtain generalized functions are
provided. With this additional insight the existence and continuity of the
mappings can be examined. In the second part of this section the
identification result is proved and sufficient conditions for the
identification mapping to be continuous are provided. A proposition about
consistent (in topology of $T^{\prime })$ nonparametric estimation that in
particular applies to functions in space $L_{1}$ completes this section.

\subsection{Results about generalized functions and existence and continuity
of mappings}

All the known results in this section are in Schwartz (1966), Gel'fand and
Shilov's monograph (vol.1 and 2, 1964) - (GS) and in Lighthill (1959)- (L);
they are listed for the convenience of the reader. The sequential approach
of Mikusinski in Antonisek et al (1973) is also referred to here. A somewhat
distinct approach to multiplication by a continuous function in the space of
generalized functions is developed at the end of this section to explain the
validity of some of the mappings in (\ref{chain}).

Definitions of generalized function spaces usually start with a topological
linear space of well-behaved "test functions", $G.$ Two most widely used
spaces are $G=D$ and $G=T$ (usually denoted $S$ in the literature). The
linear topological space of infinitely differentiable functions with finite
support $D\subset C_{\infty }(R),$ where $C_{\infty }(R)$ is the space of
all infinitely differentiable functions; convergence is defined for a
sequence of functions that are zero outside a common bounded set and
converge uniformly together with derivatives of all orders. The space $%
T\subset C_{\infty }(R)$ of test functions is defined as:%
\begin{equation*}
T=\left\{ s\in C_{\infty }(R):\left\vert \frac{d^{k}s(t)}{dt^{k}}\right\vert
=O(\left\vert t\right\vert ^{-l})\text{ as }t\rightarrow \infty ,\text{\ for
integer }k\geq 0,l>0\right\} ,
\end{equation*}%
$k=0$ corresponds to the function itself; $\left\vert \cdot \right\vert $ is
the absolute value; \ these functions go to zero faster than any power. A
sequence in $T$ converges if in every bounded region the product of $%
\left\vert t\right\vert ^{l}$ (for any $l)$ with any order derivative
converges uniformly.

A generalized function, $b,$ is defined by an equivalence class of weakly
converging sequences of test functions in \thinspace $G:$%
\begin{equation*}
b=\left\{ \left\{ b_{n}\right\} :b_{n}\in G,\text{ such that for any }s\in G,%
\text{ }\underset{n\rightarrow \infty }{\lim }\int
b_{n}(t)s(t)dt=(b,s)<\infty \right\} .
\end{equation*}%
An alternative equivalent definition is that $b$ is a linear continuous
functional on $G$ with values defined by $(b,s)^{2}\footnotetext[2]{%
As a referee pointed out this is the more commonly used definition of \
generalized function; the one above (used by S) is a necessary and
sufficient condition and thus represents an equivalent characterization.}.$\
The linear topological space of generalized functions is denoted $G^{\prime
};$ the topology is that of convergence of values of functionals for any
converging sequence of test functions (weak topology); $G^{\prime }$ is
complete in that topology. For $G=D$ or $T$ the spaces are $D^{\prime }$ and 
$T^{\prime },$ correspondingly. It is easily established that $D\subset T;$ $%
T^{\prime }\subset D^{\prime }$ and that $D^{\prime }$ has a weaker topology
than $T^{\prime },$ meaning that any sequence that converges in $T^{\prime }$
converges in $D^{\prime },$ but there are sequences that converge in $%
D^{\prime },$ but not in $T^{\prime }.$ The space $T^{\prime }$ is also
called the space of tempered distributions.

Any generalized function $b$ in $T^{\prime }$ or $\ D^{\prime }$ is (weakly)
infinitely differentiable: the generalized function $b^{(k)}$ is the $k-th$
order generalized derivative defined by $(b^{(k)},s)=\left( -1\right)
^{k}(b,s^{(k)}).$ The differentiation operator is continuous in these
spaces. For any probability distribution function $F$ on $R^{k}$ the density
function exists as a generalized function (see e.g., Zinde-Walsh, 2008) and
continuously depends on the distribution function, thus the generalized
derivative of $F$, generalized density function $f,$ is in $T^{\prime }.$

Any locally summable (integrable on any bounded set) function $b(t)$ defines
a generalized function $b$ in $D^{\prime }$ by 
\begin{equation}
(b,s)=\int b(t)s(t)dt;  \label{reg}
\end{equation}%
any such function that additionally satisfies (\ref{cond}) similarly by (\ref%
{reg}) defines a generalized function $b$ in $T^{\prime }.$ A distinction
between functions in the ordinary sense (a pointwise mapping from the domain
of definition into the reals or complex numbers) and generalized functions
is that generalized functions are not defined pointwise. Generalized
functions defined via (\ref{reg}) by ordinary functions $b(t)$ are called
regular functions; we can refer to them as ordinary regular functions in $%
G^{\prime }.$\ The functions $F,g,W_{y},W_{xy}$ are ordinary regular
functions in $T^{\prime }$ (and thus in $D^{\prime })$ if they satisfy 
Assumption 1'.

If a generalized function $b$ is such that a representation (\ref{reg}) does
not hold, it is said that $b$ is singular, so any $b\in G^{\prime }$ is
either regular or singular. A\ well-known singular generalized function is
the $\delta -$function: $\delta :$ $(\delta ,s)=s(0).$ Any generalized
function in $D^{\prime }$ or $T^{\prime }$ is a generalized finite order
derivative of a continuous function. An ordinary function that defines a
generalized function is regular if it integrates to a continuous function
and singular otherwise. For example, the ordinary function $b(t)=$ $%
\left\vert t\right\vert ^{-\frac{3}{2}}$ defines a singular generalized
function; it does not integrate to a continuous function; it does not
satisfy (\ref{reg}), in fact (see GS, v.1, p.51)\ it defines a generalized
function by 
\begin{equation}
(b,s)=\int_{0}^{\infty }t^{-\frac{3}{2}}\left\{ s(t)+s(-t)-2s(0)\right\} dt.
\label{E2}
\end{equation}

No special treatment is needed to consider complex-valued generalized
functions; all the same properties hold. For $s\in T$ or $D$ Fourier
transform $Ft(s)=\int s(t)e^{it\zeta }dt$ exists and is in $T.$ For $\ b\in
T^{\prime }:(Ft(b),s)=(b,Ft(s)),$ so $Ft(b)\in T^{\prime }.$ Fourier
transform defines a continuous and continuously invertible linear operator
in $T^{\prime }$ (but not for $D^{\prime })$. Thus Fourier transforms of $%
W_{y},W_{xy},g,$ and of the generalized derivative, $f,$ of $F$ exist in $%
T^{\prime }$ and their inverse Fourier transforms coincide with the original
functions. Since all the functions are differentiable as generalized
functions $\overset{\cdot }{\gamma }$ exists in $T^{\prime }.$ By Assumption
2 the characteristic function $\phi $ is continuous, as is its derivative, $%
\overset{\cdot }{\phi ;}$ they are regular ordinary functions in $T^{\prime
}.$

Since $G^{\prime }$ does not have a multiplicative structure, products and
convolutions can be defined for specific pairs only and generally exist only
for special classes. The product between a generalized function in $%
T^{\prime }$ and a function from $C_{\infty }$ with all derivatives bounded
by polynomial functions exists. This class of multipliers is denoted by $%
\mathcal{O}_{M}.$ Convolution of Fourier transforms of generalized functions
with Fourier transforms of functions from $\mathcal{O}_{M}$ exists and the
convolution theorem applies. Products and convolutions may exist for other
specific pairs of generalized functions. When such convolutions and products
of their Fourier transforms exist as generalized functions the convolution
theorem similarly applies to such pairs.

\textbf{Convolution Theorem.} If for $b_{1},b_{2}\in T^{\prime },$
convolution $b_{1}\ast b_{2}\in T^{\prime },$ product $Ft(b_{1})\cdot
Ft(b_{2})\in T^{\prime },$ then $Ft(b_{1}\ast b_{2})=Ft(b_{1})\cdot
Ft(b_{2}).$

The proof of this theorem uses exactly the same sequential argument as in
Antonisek et al (1973), where it utilized the specific delta-convergent
sequences; the only difference here is that the argument can be applied to
any sequence in the equivalence class that defines every given generalized
function.

To consider the product of a generalized function with a continuous function
that may not be infinitely differentiable, the property that the product
does not depend on the sequence that defines the generalized function has to
be made a requirement. We thus say that $ab$ for $b\in G^{\prime }$ and
continuous $a$ is defined in $G^{\prime }$ if for any sequence $b_{n}$ from
the equivalence class of $b$ there exists a sequence $(ab)_{n}$ in $\ G$
such that for any $\psi \in G$%
\begin{equation}
\lim \int a(x)b_{n}(x)\psi (x)dx\text{ exists and equals }\lim \int \left(
ab\right) _{n}(x)\psi (x)dx.  \label{prod}
\end{equation}

Denote by $0_{n}$ a zero-convergent sequence that belongs to the equivalence
class defining the function that is identically zero in $G^{\prime }$.

\textbf{Proposition 1}\textit{\ For the product }$ab$\textit{\ between a
continuous function }$a$ and $b\in G^{\prime }$ \textit{to be defined in }$%
G^{\prime }$\textit{\ it is necessary and sufficient that (i) }$\left( \ref%
{prod}\right) $\textit{\ hold for some sequence }$\tilde{b}_{n}$\textit{\ in
the class that defines }$b$\textit{\ and (ii) for any zero-convergent
sequence, }$0_{n}(x),$%
\begin{equation}
\lim \int a(x)0_{n}(x)\psi (x)dx=0.  \label{0converg}
\end{equation}

Proof. \ 

Any sequence $b_{n}$ differs from a specific $\tilde{b}_{n}$ by a
zero-convergent sequence.$\blacksquare $

Here we consider functions that stem from the model assumptions.
Additionally, \ we distinguish the following cases.

Case 1. Support of $\gamma $ is a bounded set: $\bar{\zeta}<\infty .$

Case 2. The function $\phi ^{-1}$\textit{\ satisfies (\ref{cond}).}

Case 3. The function $\phi \in \mathcal{O}_{M}.$

\textbf{Lemma 1} \textit{Under Assumptions 1-3 }

\textit{(i) the products }$\gamma \phi $\textit{\ and }$\gamma \dot{\phi}$ 
\textit{are defined in }$T^{\prime }$\textit{\ and in }$D^{\prime };$\textit{%
\ }

\textit{(ii) for }$\tilde{\phi}^{-1}=\phi ^{-1}(\zeta )I(\left\vert \zeta
\right\vert <\bar{\zeta})$ \textit{the product }$(\gamma \phi )\cdot \tilde{%
\phi}^{-1}$\textit{\ is always defined in }$D^{\prime };$

\textit{(iii) if either case 1 applies or both cases 2 and 3 apply the
product }$(\gamma \phi )\cdot \tilde{\phi}^{-1}$\textit{\ is defined in }$%
T^{\prime }$\textit{; }

(iv) \textit{if neither case 1 nor case 2 applies the product may not be
defined in }$T^{\prime }.$

Proof. See Appendix.

From Lemma 1 existence of products to justify the convolution theorem and
thus equations (\ref{Fy},\ref{Fxy}) follows. The mapping ($\ref{S})$
involves solving equations (\ref{Fy},\ref{Fxy}) for the unknown functions\
and requires multiplication by $\phi ^{-1};$ as one can see from Lemma 1
existence of such products in $T^{\prime }$ is not guaranteed.

\subsection{The nonparametric identification theorem}

This section contains two results. The first is Theorem 1 that\ proves the
existence of the identification mapping $M^{\ast }$ under Assumptions 1-3.
It differs from the statement in (S, Theorem 1) in three ways: first,
Assumption 1 (and even the more restrictive Assumption 1') of this paper is
more general; second, it does not rely on decomposition of generalized
functions$^{3}$\footnotetext[3]{%
There is no known decomposition in the space of generalized functions\ into
generalized functions corresponding to ordinary functions and to singular
functions, claimed in (S); the pointwise argument provided in (S, 2007,
Supplementary Material) is incorrect (see Appendix B).}; third, it provides
the condition under which the mapping is obtained via operations in the
space $T^{\prime }$ and discusses the continuity of the identification
mapping. The second result is Proposition 2 that shows that when continuity
holds, consistent (in the topology of $T^{\prime })$ plug-in non-parametric
estimation of the regression function is possible, e.g. for functions in
space $L_{1}$.

\textbf{Theorem 1 }\textit{For functions satisfying model assumptions and
Assumptions 1-3 the mapping }$M^{\ast }$ \textit{in (\ref{M}) exists and
provides identification for }$g$\textit{; if conditions of (iii) of Lemma 1
are satisfied the mapping is defined via operations in }$T^{\prime }$\textit{%
; it can be discontinuous under condition (iv) of Lemma 1.}

Proof. See Appendix.

The implication of this Theorem is that the identification result holds
under the general assumptions 1-3. If $\phi $ is too thin-tailed, however,
the mapping whereby the identification is achieved may not be continuous:
this point is illustrated by the example in the proof of Theorem 1 where
high frequency components $b_{n}$ are magnified by multiplication with $\phi
^{-1}$ from a thin-tailed distribution; this produces inverse Fourier
Transforms that diverge.

Continuity requires that the mapping $M$ given by the model assumptions be
continuous in $T^{\prime }.$ Continuity in $T^{\prime }$ allows for a
consistent (in $T^{\prime })$ plug-in estimator; the following Proposition
provides sufficient conditions. Denote by $\rightarrow _{T^{\prime }}$
convergence in topology of $T^{\prime }.$ Following Gel'fand and Vilenkin
(1964) we define a random generalized function as the random continuous
functional on the space of test functions.

\textbf{Proposition 2 }\textit{(a) Under the conditions of Theorem 1} 
\textit{suppose that }$W_{yn},W_{xyn}$ \textit{are random generalized
functions (estimators) that satisfy model assumptions and Assumptions 1-3
together with some (unknown) functions }$g_{n},F_{n}$\textit{; condition
(iii) of Lemma 1 is satisfied; }$\phi _{n}\in \mathcal{O}_{M}$\textit{; for }%
$\varepsilon _{yn}=Ft(W_{yn}),$ $\varepsilon _{xyn}=Ft(W_{xyn})$ \textit{%
assume that the Fourier transforms satisfy: }$\varepsilon _{yn}(\zeta )$ 
\textit{is continuous and non-zero a.e. on }supp($\gamma )$ \textit{and }$%
i\varepsilon _{xyn}(\zeta )-\dot{\varepsilon}_{yn}(\zeta )$ \textit{is
continuous and that\ }%
\begin{eqnarray}
\Pr (\varepsilon _{yn}(\zeta ) &\rightarrow &_{T^{\prime }}\varepsilon
_{y}(\zeta ))\rightarrow 1;  \label{converg} \\
\Pr (\varepsilon _{xyn}(\zeta ) &\rightarrow &_{T^{\prime }}\varepsilon
_{xy}(\zeta ))\rightarrow 1,  \notag
\end{eqnarray}%
\textit{then it is possible to find a sequence }$g_{n}(x)$\textit{\ such
that }$\Pr (g_{n}(x)\rightarrow _{T^{\prime }}g(x))\rightarrow 1.$

\textit{(b) Suppose that the function }$g(x)\in L_{1}.$\textit{\ }Then there
exists a sequence of step function estimators, $g_{n},$ such that 
\begin{equation*}
\Pr (g_{n}(x)\rightarrow _{T^{\prime }}g(x))\rightarrow 1.
\end{equation*}

Proof. See Appendix.

Convergence of the estimators is in the weak topology of space $T^{\prime }$
of generalized functions, not in $L_{1}.$ If $g_{n}(x)\rightarrow
_{T^{\prime }}g(x)$ and $g(x)$ is a continuous function then there is
pointwise convergence and uniform convergence on bounded sets. 

\section{ Semiparametric specification and identification}

Semiparametric models with measurement error were examined for polynomial
regression functions by Hausman et al (1991), for regression function in the 
$L_{1}$ space \ by Wang and Hsiao (1995, 2009). (S) significantly widened
the classes of semiparametric models with errors-in-variables where
identification can be achieved via moment conditions by utilizing
generalized functions, but did not explicitly characterize the class of
functions which she considered: verifying moment conditions of (S,
Assumption 6) is needed. In contrast, the class of parametric functions is
characterized directly in Assumptions 5 and 6 of this paper; the assumptions
give sufficient conditions for identification via moments. The results in
(S) rely on existence of a moment generating function for the measurement
error; this restriction in not imposed here. In this paper as well as in (S)
some moment conditions involve limits for sequences of weighting functions;
the limits are explicitly given here.

\textbf{Assumption 4.} The function $g(x^{\ast })$ is in a parametric class
of locally integrable functions $g(x^{\ast },\theta )$ where $\theta \in
\Theta ;$ $\Theta $ is an open set in $R^{m};$ for some $\theta ^{\ast }\in
\Theta $ model assumptions and (\ref{W1}) hold.

Denote all the Fourier transforms of the parametric functions in the model
assumptions as $\gamma (\theta );\varepsilon _{y}(\theta );\varepsilon
_{xy}(\theta ).$ The following assumption restricts the generalized function 
$\gamma (\theta )$ to have no more than a finite number of special points: $%
\Delta $ points of singularity and $J$ of "jump" discontinuity in some
region $\left\vert \zeta \right\vert <\bar{\zeta}<\infty $. Notation $\left[
x\right] $ is for integer part of $x;$ $\delta (\zeta -a)$ denotes a shilted 
$\delta -function:$ ($\delta (\zeta -a),\psi )=\psi (a)$ for $\psi \in G.$ 

\textbf{Assumption 5.} The Fourier transform, $\gamma (\theta ),$ of the
real function $g(x^{\ast },\theta )$ in the region $\left\vert \zeta
\right\vert <\bar{\zeta}<\infty $ that belongs to its support (and may
coincide with it) can be represented as 
\begin{equation}
\gamma (\theta )=\gamma _{o}(\theta )+\gamma _{s}(\theta ),  \label{a5}
\end{equation}%
where

(i) if $\Delta =0,$ $\gamma _{s}(\theta )\equiv 0$ $;$ if $\Delta \geq 1$  
\begin{eqnarray}
\gamma _{s}(\theta ) &=&2\pi \dsum\limits_{l=0}^{L}\gamma _{s_{l}}(\theta ),%
\text{ where }L=\left[ \frac{\Delta }{2}\right] \text{ and} \\
\gamma _{s_{l}}(\theta ) &=&\dsum\limits_{k=0}^{\bar{k}_{l}}\left( \gamma
_{k}(s_{l},\theta )\delta ^{(k)}(\zeta -s_{l})+\overline{\gamma
_{k}(s_{l},\theta )}\delta ^{(k)}(\zeta +s_{l})\right) ;\text{ for }%
l=0,1,...L;  \notag
\end{eqnarray}

\ (ii) $\gamma _{o}(\theta )\equiv \gamma _{o}(\zeta ,\theta )$ is defined
by a locally integrable function of $\zeta $ continuous except possibly in a
finite number of points and such that its generalized derivative, $\overset{%
\cdot }{\gamma }_{o}(\theta ),$ is of the form%
\begin{equation*}
\overset{\cdot }{\gamma }_{o}(\theta )=\overset{\cdot }{\gamma }_{oo}(\theta
)+\overset{\cdot }{\gamma }_{os}(\theta ),
\end{equation*}%
where if $J=0,$ then $\overset{\cdot }{\gamma }_{os}(\theta )=0,$ and if $%
J>0,$ then for points $b_{j},$ $j=1,...\left[ \frac{J}{2}\right] ,$ $\overset%
{\cdot }{\gamma }_{os}(b_{j},\theta )=$ 
\begin{equation*}
\gamma _{os0}(0,\theta )\delta (\zeta )I(J\text{ is odd})+\Sigma _{j=1}^{%
\left[ \frac{J}{2}\right] }\left( \gamma _{osj}(b_{j},\theta )\delta (\zeta
-b_{j})+\overline{\gamma _{osj}(b_{j},\theta )}\delta (\zeta +b_{j})\right) ,
\end{equation*}%
$\overset{\cdot }{\gamma }_{oo}(\theta )\equiv \overset{\cdot }{\gamma }%
_{oo}(\zeta ,\theta )$ is an ordinary locally integrable function continuous
except possibly in a finite number of points;

(iii) $\gamma _{o}(\zeta ,\theta )\neq 0$ except possibly for a finite
number of points in $(-\bar{\zeta},\bar{\zeta});$

(iv) At any non-zero singularity point: $s_{l}\neq 0,$ $\gamma _{o}(\zeta
,\theta )$ is continuous and non-zero. \ \ \ \ \ \ \ \ \ \ 

Under Assumptions 1 and 5 $g$ could be in $L_{1},$ or a sum of a function
from $L_{1}$ and a polynomial (singularity point $\zeta _{0}=0)$ and also
possibly a periodic function, e.g. $\sin \left( \cdot \right) $ or $\cos
\left( \cdot \right) $ with singularities at some points $\pm s,$ $s\neq 0$.
Here the parameters, $\gamma _{\cdot }(\cdot ,\theta ),$ are allowed to take
complex values, otherwise one would need to be more specific about the
functions with singular Fourier transforms; since the functions are assumed
known it is easy in each specific case to separate out the imaginary parts
as in the case of polynomials.

Assumption 5 permits to write moment conditions; however, to get a
sufficient condition for identification of all parameters additionally the
following Assumption 6 is made.

If $\Delta >0$ define the matrices $\Gamma _{y}(s_{l},\theta )$ and $\Gamma
_{xy}(s_{l},\theta )$ for each $s_{l}\geq 0$ (similarly to (S) for the case $%
s_{l}=0$ ) by their elements:%
\begin{eqnarray*}
\Gamma _{y,i+1,k+1}(s_{l},\theta ) &=&\left( 
\begin{array}{c}
k+i \\ 
i%
\end{array}%
\right) \gamma _{k+i}(s_{l},\theta )I(k+i\leq \bar{k}_{l}); \\
\Gamma _{xy,i+1,k+1}(s_{l},\theta ) &=&\left( 
\begin{array}{c}
k+i+1 \\ 
i+1%
\end{array}%
\right) \gamma _{k+i}(s_{l},\theta )I(k+i\leq \bar{k}_{l}), \\
i,k &=&0,1,...\bar{k}_{l}.
\end{eqnarray*}%
Denote by $\left\{ A\right\} _{11}$ the first matrix element of a matrix $A.$

\textbf{Assumption 6.} The function $\gamma $ satisfies Assumption 5.
Additionally all $\gamma _{o}(\zeta ,\theta ),\overset{\cdot }{\gamma }%
_{oo}(\zeta ,\theta ),$ $\gamma _{s_{l}}(\theta )$ are continuously
differentiable with respect to the parameter, $\theta ,$ in some
neighborhood of $\theta ^{\ast }.$ The $m\times 1$ parameter vector can be
partitioned as $\theta ^{T}=[\theta _{I}^{T};\theta _{II}^{T}].$ For any
component, $\theta _{i},$ of $m_{I}\times 1$ vector $\theta _{I}$ (where $%
m\geq m_{I}\geq 0)$ either%
\begin{equation}
\gamma _{o}(\zeta ,\theta ^{\ast })\frac{\partial }{\partial \theta _{i}}%
\overset{\cdot }{\gamma }_{oo}(-\zeta ,\theta )|_{\theta ^{\ast }}+\overset{%
\cdot }{\gamma }_{oo}(\zeta ,\theta ^{\ast })\frac{\partial }{\partial
\theta _{i}}\gamma _{o}(-\zeta ,\theta )|_{\theta ^{\ast }}\neq 0\text{ }
\label{ordid}
\end{equation}%
a.e., or if (\ref{ordid}) does not hold for some $i^{\ast }$, then $\frac{%
\partial }{\partial \theta _{i^{\ast }}}\gamma _{o}(-\zeta ,\theta
)|_{\theta ^{\ast }}\neq 0.$ If $m_{II}>0$ the matrix that stacks for all $%
s_{l},$ $l\geq 0$ matrices 
\begin{equation*}
\left( 
\begin{array}{c}
\frac{\partial }{\partial \theta _{II}^{T}}\left[ \Gamma _{y}(s_{l},\theta )%
\right] ^{-1}|_{\theta ^{\ast }}\Gamma _{y}(s_{l},\theta ^{\ast })+\frac{%
\partial }{\partial \theta _{II}^{T}}\left[ \Gamma _{xy}(s_{l},\theta )%
\right] ^{-1}|_{\theta ^{\ast }}\Gamma _{xy}(s_{l},\theta ^{\ast }) \\ 
\frac{\partial }{\partial \theta _{II}^{T}}\left\{ \left[ \Gamma
_{y}(s_{l},\theta )\right] ^{-1}|_{\theta ^{\ast }}\Gamma _{y}(s_{l},\theta
^{\ast })\right\} _{11}|_{\theta ^{\ast }}%
\end{array}%
\right)
\end{equation*}%
is of rank $m_{II}.$

By checking we can see that all the examples provided in (S) satisfy
assumptions 5 and 6 here and thus sufficient conditions for identification
hold. If the same parameters enter into both the ordinary and singular parts
(S, assumption 6) may be violated, even though identification is possible
and the results of this paper hold.

Additional assumptions 7 and 8 below are needed.$^{4}$\footnotetext[4]{%
Assumptions 7 and 8(ii) are also implicit in the proofs in (S).}

\textbf{Assumption 7.} The density function $p(z)$ exists and is positive.

\textbf{Assumption 8.} The characteristic function of measurement error, $%
\phi (\zeta ),$ is such that (i) $\phi (\zeta )\neq 0$ for $\left\vert \zeta
\right\vert <\bar{\zeta}$ where it is continuously differentiable; (ii) it
is $\bar{k}_{l}$ times continuously differentiable at every $s_{l}.$

Theorem 2 below establishes that moment conditions for the parameters $%
\theta $ of $\gamma (\zeta ,\theta )$ hold and Theorem 3 that the
assumptions are sufficient for identification. The notation $\func{Re}(x)$
refers to the real part of a complex $x.$

\textbf{Theorem 2.}\textit{\ Under model assumption and  Assumptions 1', 4,
7, 8}

\textit{(i) if Assumption 5 (i,ii) holds there exist real functions }$%
r_{y}(z,\theta ),r_{xy}(z,\theta )$\textit{\ such that the moment}%
\begin{equation}
E\left( \frac{Yr_{xy}(z,\theta )+XYr_{y}(z,\theta )}{p(z)}\right) 
\label{gammaoshape}
\end{equation}%
\textit{exists for }$\theta $\textit{\ in some neighborhood of }$\theta
^{\ast }$\textit{\ and equals zero for }$\theta =\theta ^{\ast };$

\textit{(ii) if 5(i-iii) holds there are functions }$r_{y1n}(z,\theta )$%
\textit{\ such that} 
\begin{equation}
\underset{n\rightarrow \infty }{\lim }E\left( \frac{Yr_{y1n}(z,\theta )}{p(z)%
}-1\right)  \label{gammaoscale}
\end{equation}%
\textit{exists for }$\theta $\textit{\ in some neighborhood of }$\theta
^{\ast }$\textit{\ and equals zero for }$\theta =\theta ^{\ast };$

\textit{(iii) If }$\Delta >0$\textit{\ and 5(i-ii) hold then for each }$%
s_{l}\geq 0$\textit{\ there exist vector functions }$r_{ysl}(z,\theta ),$%
\textit{\ }$r_{ysl,n}(z,\theta ),r_{xysl}(z,\theta ),r_{xysl,n}(z,\theta ),$%
\textit{\ and a diagonal invertible matrix }$M_{l}$\textit{\ such that}%
\begin{eqnarray}
&&\underset{n\rightarrow \infty }{\lim }\func{Re}[\Gamma
_{y}^{-1}(s_{l},\theta )M_{l}^{-1}E\left( \frac{Yr_{ys,l,n}(z,\theta )}{p(z)}%
\right)  \label{gammasshape} \\
&&+\Gamma _{xy}^{-1}(s_{l},\theta )M_{l}^{-1}E\left( \frac{%
XYr_{xys,l,n}(z,\theta )}{p(z)}\right) ]  \notag
\end{eqnarray}%
\textit{exists for }$\theta $\textit{\ in some neighborhood of }$\theta
^{\ast }$\textit{and equals zero for }$\theta =\theta ^{\ast };$

\textit{(iv) If }$\Delta >0$\textit{\ and 5(i-iv) hold then for each }$%
s_{l}\geq 0$\textit{\ there exist functions }$r_{ysl,1,n}(z,\theta ),$%
\textit{\ }$r_{yslo,1,n}(z,\theta )$\textit{\ such that for }$s_{0}=0$%
\textit{\ }%
\begin{equation}
\underset{n\rightarrow \infty }{\lim }E\left( \frac{Yr_{ys0,1,n}(z,\theta )}{%
p(z)}-1\right)  \label{gamma0scale}
\end{equation}%
\textit{\ and }%
\begin{equation}
\underset{n\rightarrow \infty }{\lim }\func{Re}E\left( \frac{Y\left(
r_{ysl,1,n}(z,\theta )-r_{yslo,1,n}(z,\theta )\right) }{p(z)}\right)
\label{gammasscale}
\end{equation}%
\textit{exist for }$\theta $\textit{\ in some neighborhood of }$\theta
^{\ast }$\textit{and equal zero for }$\theta =\theta ^{\ast }.$

Proof. See Appendix. The functions $r_{\cdot }(z,\theta )$ and matrices $%
M_{l}$ are provided there.

Some of the moment conditions can be redundant. Different sets of weighting
functions could be appropriate; similarly to reasoning in (S) the weighting
functions are designed in a way that isolates different components of the $%
\gamma $ function: the ones in (i) are for the ordinary function component
and are supplemented by moments in (ii) for the case of a scale multiple for
the ordinary component, the ones in (iii) are for the coefficients of the
singular part with (iv) for the possible scale factor at each singularity.
If only (\ref{gammaoshape}) applies then the weighting functions proposed in
(S) can be used, but for the other components the weights proposed here
solve the problem without additional requirements that moment generating
function for errors exist and avoid the problematic function $\mu $ in (S,
Definition 2): $\mu (0)$ and any derivatives of $\mu $ at 0 are zero (see
Appendix B).

Define by $EQ(\theta )$ the vector with components provided by the stacked
expressions (whichever are defined) from (\ref{gammaoshape}, \ref%
{gammaoscale}, \ref{gammasshape}, \ref{gammasscale}).

\textbf{Theorem 3.}\textit{\ Under the conditions of Theorem 2 and
Assumption 6 the functions }$r_{\cdot }(z,\theta )$\textit{\ can be selected
in such a way that the matrix }$\frac{\partial }{\partial \theta }EQ(\theta
^{\ast })$\textit{\ exists and has rank }$m$\textit{.}

Proof. See Appendix.

Theorem 3 provides sufficient conditions under which the equations $%
EQ(\theta )=0$ fully identify the parameter\ vector $\theta ^{\ast }.$

\section{Appendix A}

\subsection{\protect\bigskip\ Proofs}

\textbf{Proof of Lemma 1.}

(i) By model assumption and Assumption 1 the convolutions in (\ref{W1})
provide elements in $T^{\prime };$ this implies that any of the sequential
definitions of convolution in Kamiski (1982) hold, therefore by his theorem
9, the "exchange formula" implies that the products for some sequences in
the equivalence classes defining the $Ft^{\prime }s$ exist. By Proposition 1
it follows that since $\phi ,\overset{\cdot }{\phi }\in T^{\prime },$ (\ref%
{0converg}) holds for continuous functions $\phi $ and $\overset{\cdot }{%
\phi }\ $then $\gamma \phi \in T^{\prime },$ $\overset{\cdot }{\gamma }\phi
\in T^{\prime }$ and additionally (by applying the product rule to (\ref{Fy},%
\ref{Fxy})) $\gamma \dot{\phi}\in T^{\prime }.$ Since $T^{\prime }\subset
D^{\prime }$, the products are defined in $D^{\prime }$ as well.

(ii) Now consider a sequence $(\gamma \phi )_{n}$ defined as follows: select
some sequence $\tilde{\gamma}_{n}$ for $\gamma $ from $D;$ then each $\tilde{%
\gamma}_{n}$ has finite support; for a sequence of numbers $\varepsilon
_{n}\rightarrow 0$ select $\tilde{\phi}_{n}$ in $D$ such that $\left\vert 
\tilde{\phi}_{n}-\phi \right\vert <\frac{\varepsilon _{n}}{\sup \left\vert 
\tilde{\gamma}_{n}\phi ^{-1}\right\vert }$ on compact support of $\gamma
_{n}.$ Then for the sequence $(\gamma \phi )_{n}=\tilde{\gamma}_{n}\tilde{%
\phi}_{n}$ and any $\psi \in D$%
\begin{equation*}
\int \tilde{\gamma}_{n}\tilde{\phi}_{n}\phi ^{-1}\psi =\int \tilde{\gamma}%
_{n}\psi +\int \tilde{\gamma}_{n}(\tilde{\phi}_{n}-\phi )\phi ^{-1}\psi
\rightarrow \int \gamma \psi .
\end{equation*}%
Now we check that (\ref{0converg}) holds for $a=\phi ^{-1}.$ In $D$ support
of any $\psi $ is bounded, on that compact set $\phi ^{-1}$ is bounded thus (%
\ref{0converg}) will hold and the product is defined in $D^{\prime }$.

(iii) For Case 1 the product with $\phi ^{-1}(\zeta )I(\left\vert \zeta
\right\vert <\bar{\zeta})$ is similarly to (ii) defined in $T^{\prime }$
since it is sufficient to consider $\psi \in T$ with bounded support
(containing support of $\gamma ).$ If cases 2 and 3 hold it is
straightforward to verify that the function $\phi ^{-1}$ is in $\mathcal{O}%
_{M}$, thus the product is defined (continuously) in $T^{\prime }$.

(iv) We construct a counterexample. The function $\phi (x)=e^{-x^{2}}$ does
not belong to either case 1 or case 2. The product of function $b(x)\equiv 0$%
\ and $\phi (x)^{-1}$\ does not exist in\textit{\ }$T^{\prime }$\textit{. }%
Consider $b_{n}(x)=$

\begin{equation}
\left\{ 
\begin{array}{cc}
e^{-n} & \text{if }n-\frac{1}{n}<x<n+\frac{1}{n}; \\ 
0\leq b_{n}(x)\leq e^{-n} & \text{if }n-\frac{2}{n}<x<n+\frac{2}{n}; \\ 
0 & \text{otherwise.}%
\end{array}%
\right.  \label{bn}
\end{equation}%
This $b_{n}(x)$ converges to $b(x)\equiv 0$ in $T^{\prime }$. Indeed for any 
$\psi \in T$ 
\begin{equation*}
\int b_{n}(x)\psi (x)dx=\int_{n-2/n}^{n+2/n}b_{n}(x)\psi (x)dx\rightarrow 0.
\end{equation*}

\bigskip But the sequence $b_{n}(x)\tilde{\phi}(x)^{-1}$\ does not converge
in the space $T^{\prime }$ of tempered distributions. Indeed if it did then $%
\int b_{n}\tilde{\phi}\psi $ would converge for any $\psi \in T.$ But for $%
\psi \in T$ such that $\psi (x)=\exp (-\left\vert x\right\vert )$ for, e.g. $%
\left\vert x\right\vert >1$ 
\begin{eqnarray*}
\int_{n-2/n}^{n+2/n}b_{n}(x)e^{x^{2}}\psi (x)dx &\geq
&e^{-n}\int_{n-1/n}^{n+1/n}e^{x^{2}-x}dx \\
&\geq &\frac{2}{n}e^{-2n+\left( n-1/n\right) ^{2}}.
\end{eqnarray*}%
This diverges.$\blacksquare $

\textbf{Proof of Theorem 1.}

The proof makes use of different spaces of generalized functions and
exploits relations between them. It proceeds in two parts.

First in part one, it is shown that from equations (\ref{Fy},\ref{Fxy}) the
continuous function $\varkappa =\dot{\phi}\phi ^{-1}$ can be uniquely
pointwise determined on the interval $[-\bar{\zeta},\bar{\zeta}]$ (which is
in the support of $\gamma $ and consequently of $\varepsilon _{1})$; this
requires additionally considering the generalized functions spaces, $%
D^{\prime }$ and also $D_{0}(U)^{\prime }$ which is defined on the space of
test functions that are continuous with support contained in $U.$ The
function $\phi $ is uniquely defined on the interval $(-\bar{\zeta},\bar{%
\zeta})$ as the solution of the corresponding differential equation that
satisfies the condition $\phi (0)=1.$ Define $\tilde{\phi}=\phi I(\left\vert
\zeta \right\vert <\bar{\zeta})$; define $\tilde{\phi}^{-1}$ to equal $\phi
^{-1}I(\left\vert \zeta \right\vert <\bar{\zeta})$. Of course, when $\bar{%
\zeta}=\infty ,$ $\tilde{\phi}=\phi $ and $\tilde{\phi}^{-1}=\phi ^{-1}$\ on 
$R.$

Next in part two, $\gamma $ is defined as $\varepsilon _{y}\tilde{\phi}^{-1}$%
. By Lemma 1 this product can always be uniquely defined as a generalized
function in $D^{\prime };$ by construction $\gamma $ defines a generalized
function in $T^{\prime }\subset D^{\prime };$ this provides the required
mapping $M^{\ast }$ by applying inverse Fourier Transform to $\gamma $. If $%
\ $condition (iii) of Lemma 1 applies the product\textit{\ }$\varepsilon _{y}%
\tilde{\phi}^{-1}$ is defined in $T^{\prime }$; in this case all the
mappings that define the mapping $M^{\ast }$ are defined in $T^{\prime }$.
The proof concludes with an example that demonstrates that the mapping can
be discontinuous if (iv) of Lemma 1 applies.

Part one. Consider the space of generalized functions $D^{\prime }.$ By
Assumption 2 $\phi $ is non-zero and continuously differentiable, then by
differentiating (\ref{Fy}), substituting (\ref{Fxy}) and making use of Lemma
1 to multiply by $\phi ^{-1}$ in $D^{\prime }$ we get that the generalized
function%
\begin{equation*}
\varepsilon _{y}\phi ^{-1}\dot{\phi}-(\dot{\varepsilon}_{y}-i\varepsilon
_{xy})
\end{equation*}%
equals zero in the sense of generalized function in $D^{\prime }.$ Denoting $%
\varkappa =\dot{\phi}\phi ^{-1}$ we can characterize $\varkappa $ as a
continuous at every point (by Assumption 2) function in $D^{\prime }$ that
satisfies the equation 
\begin{equation}
\varepsilon _{y}\varkappa -(\dot{\varepsilon}_{y}-i\varepsilon _{xy})=0.
\label{f}
\end{equation}%
If (\ref{f}) holds in $D^{\prime },$ it holds also for any test functions
with support limited to $U:$ $\psi \in D(U)\subset D,$ and thus holds in any 
$D(U)^{\prime }.$

We show that the function $\varkappa $ is uniquely determined in the class
of continuous functions on on $[-\bar{\zeta},\bar{\zeta}]$ by (\ref{f})
holding in $D(U)^{\prime }$\ for any interval $U\subset (-\bar{\zeta},\bar{%
\zeta}).$ Proof is by contradiction. Suppose that there are two distinct
continuous functions on $[-\bar{\zeta},\bar{\zeta}],$ $\varkappa _{1}\neq
\varkappa _{2}$ that satisfy (\ref{f}), then $\varkappa _{1}(\bar{x})\neq
\varkappa _{2}(\bar{x})$ for some $\bar{x}\in (-\bar{\zeta},\bar{\zeta});$
by continuity $\varkappa _{1}\neq \varkappa _{2}$ everywhere for some
interval $U\subset (-\bar{\zeta},\bar{\zeta}).$ Consider now $D(U)^{\prime
}; $ we can write 
\begin{equation*}
(\varepsilon _{y}(\varkappa _{1}-\varkappa _{2}),\psi )=0
\end{equation*}%
for any $\psi \in D(U).$ A generalized function that is zero for all $\psi
\in D(U)$ coincides with the ordinary zero function on $U$ and is also zero
for all $\psi \in D_{0}(U)$, where $D_{0}$ denotes the space of continuous
test functions. For the space of test function $D_{0}(U)$ multiplication by
continuous $(\varkappa _{1}-\varkappa _{2})\neq 0$ is an isomorphism. Then
from (\ref{f}) we can write%
\begin{equation*}
0=(\left[ \varepsilon _{y}(\varkappa _{1}-\varkappa _{2})\right] ,\psi
)=(\varepsilon _{y},(\varkappa _{1}-\varkappa _{2})\psi )
\end{equation*}%
implying that $\varepsilon _{y}$ is defined and is a zero generalized
function in $D_{0}(U)^{\prime }.$ If that were so $\varepsilon _{y}$ would
be a zero generalized function in $D(U)^{\prime }$ since $D(U)\subset
D_{0}(U);$ this contradicts Assumption 2. This concludes the first part of
the proof since from $\varkappa $ the function 
\begin{equation*}
\phi (\zeta )=\exp \int_{0}^{\zeta }\varkappa (\xi )d\xi
\end{equation*}%
that solves on $[-\bar{\zeta},\bar{\zeta}]$ 
\begin{equation*}
\dot{\phi}\phi ^{-1}=\varkappa ;\phi (0)=1
\end{equation*}%
is uniquely determined on $[-\bar{\zeta},\bar{\zeta}]$ and $\tilde{\phi}$
(and $\tilde{\phi}^{-1})$ defined above are uniquely determined.

Part two.

Consider two cases.

Case 1.Part (iii) of Lemma 1 applies. Multiplying $\varepsilon _{y}(=\gamma 
\tilde{\phi})$ by $\tilde{\phi}^{-1}$ provides a tempered distribution by
Lemma 1; it is equal to $\gamma .$ The inverse Fourier Transform provides $g.
$ The theorem holds and moreover, all the operations by which the solution
was obtained were defined in $T^{\prime }$.

Case 2. The condition (iii) of Lemma 1 may not hold, so multiplication by $%
\tilde{\phi}^{-1}=\phi ^{-1}$ may not lead to a tempered distribution.
Consider now $D^{\prime };$ $T^{\prime }\subset D^{\prime }.$ Multiplication
by $\phi ^{-1}$ is a continuous operation in $D^{\prime }$; define the same
differential equations, solve to obtain $\phi $ and get via multiplication $%
(\gamma \phi )\cdot \phi ^{-1}$ in $D^{\prime }$ the function $\gamma \in
D^{\prime }$. Since $\gamma $ is the Fourier transform of $g$ (a tempered
distribution) it also belongs to $T^{\prime },$ and it is possible to
recover $g$ by an inverse Fourier Transform.

In the following example the mapping $M^{\ast }$ in (\ref{M}) is not
continuous. Define $\beta _{n}=Ft^{-1}(b_{n})$ where $b_{n}$ is defined by (%
\ref{bn}); $b_{n}\in T,$ thus $\beta _{n}\in T.$ In was shown in proof of
Lemma 1 that $b_{n}(x)$ converges to $b(x)\equiv 0$ in $T^{\prime }$.

Suppose that the model mapping $M$ in (\ref{chain}) is defined for functions
in $L_{1}$ and is continuous in $L_{1}$ (and thus in $T^{\prime }).$ Suppose
that $W_{yn}=W_{y}+\beta _{n};$ from $b_{n}\rightarrow 0$ in $T^{\prime }$
and the continuity of the Fourier Transform mapping in $T^{\prime },$ it
follows that $\beta _{n}\rightarrow 0$ and thus $W_{yn}\rightarrow W_{y}$ in 
$T^{\prime }.$ Then $\varepsilon _{yn}=\varepsilon _{y}+b_{n}.$ Suppose that 
$\phi $ is proportionate to $e^{-x^{2}}$. Then by the proof in part 1 each $%
\gamma _{n}=\varepsilon _{yn}\phi ^{-1}\in D^{\prime },$ but as a Ft of a
function in $T^{\prime }$ (even in $L_{1}$ here) is defined in $T^{\prime };$
the inverse Fourier transform, $\tilde{g}_{n}=Ft^{-1}(\gamma _{n}),$ exists.
However, $\tilde{g}_{n}$ does not converge to $g$ in $T^{\prime }.$ Indeed,
if it did so converge, then that would imply convergence $\gamma
_{n}\rightarrow \gamma $ in $T^{\prime }$, but $b_{n}(x)e^{x^{2}}$\ does not
converge in the space $T^{\prime }$ of tempered distributions as was shown
in the proof of Lemma 1. $\blacksquare $

\textbf{Proof of Proposition 2.}

(a) We establish that the mapping from $(W_{yn},W_{xyn})$ to $g_{n}$ is
continuous. Consider $\zeta \in $supp($\gamma ).$ Similarly to proof in
Theorem 1 applied to every $n$ a continuous function $\varkappa _{n}(\zeta
), $ that satisfies the equation%
\begin{equation}
\varkappa _{n}(\zeta )\varepsilon _{yn}(\zeta )+(i\varepsilon _{xyn}(\zeta )-%
\dot{\varepsilon}_{yn}(\zeta ))=0  \label{20}
\end{equation}%
in generalized functions, exists\ (defined as $\dot{\phi}_{n}\phi _{n}^{-1})$
and is unique. Moreover, from Lemma 1 it follows that the product with $%
\varkappa _{n}=\dot{\phi}_{n}\phi _{n}^{-1}\in \mathcal{O}_{M}$ always
exists. Since all functions in (\ref{20}) are continuous it represents an
equality of continuous functions and since $\varepsilon _{yn}$ is non-zero
a.e. we have \textit{\ }%
\begin{equation*}
\varkappa _{n}(\zeta )=(i\varepsilon _{xyn}(\zeta )-\dot{\varepsilon}%
_{yn}(\zeta ))(\varepsilon _{yn}(\zeta ))^{-1}.
\end{equation*}

The generalized functions

\begin{equation*}
\varkappa _{n}\varepsilon _{yn}-\varkappa \varepsilon _{y}=i(\varepsilon
_{xyn}-\varepsilon _{xy})+(\dot{\varepsilon}_{yn}-\dot{\varepsilon}_{y})%
\text{ and }\varkappa _{n}(\varepsilon _{y}-\varepsilon _{yn})
\end{equation*}%
converge to zero as generalized functions in $T^{\prime }$; as a result, so
does $(\varkappa _{n}-\varkappa )\varepsilon _{yn},$ but since this is a
continuous function this implies pointwise convergence. Suppose that on some
bounded interval $\varkappa _{n}-\varkappa $ is separated away from zero for
some subsequence $\left\{ n_{i}\right\} $, this implies then that on that
set $\varepsilon _{yn_{i}}$ converges to zero pointwise, thus the limit in $%
T^{\prime }$ (=$\varepsilon _{y})$ is zero on this interval which belongs to
support of $\gamma ,$ and thus of $\varepsilon _{y}.$ This contradiction
establishes that $\varkappa _{n}\rightarrow \varkappa $ pointwise and
uniformly on any bounded set. From the differential equation $\phi _{n}^{-1}%
\dot{\phi}_{n}=\varkappa _{n}$ with the condition $\phi _{n}(0)=1$ the
function $\phi _{n}$ is uniquely determined; and $\phi _{n}\rightarrow \phi $
where $\phi ^{-1}\dot{\phi}=\varkappa ,\phi (0)=1.$ Then also since $\phi $
is non-zero, $\phi _{n}^{-1}\rightarrow \phi ^{-1}$ pointwise; $\phi
_{n}^{-1}$ satisfies \textit{(\ref{cond})} so that $\varepsilon _{1n}\phi
_{n}^{-1}$ can be defined as a tempered distribution. Finally consider%
\begin{equation*}
\varepsilon _{yn}\phi _{n}^{-1}-\varepsilon _{y}\phi ^{-1}=\varepsilon
_{yn}(\phi _{n}^{-1}-\phi ^{-1})+(\varepsilon _{yn}-\varepsilon _{y})\phi
^{-1}.
\end{equation*}%
The continuous function $\varepsilon _{yn}(\phi _{n}^{-1}-\phi
^{-1})\rightarrow 0$; the generalized function $(\varepsilon
_{yn}-\varepsilon _{y})\phi ^{-1}\rightarrow 0$ in $T^{\prime }$ (as a
tempered distribution) $\varepsilon _{yn}\phi _{n}^{-1}$ converges to $%
\gamma $ in $T^{\prime },$ and its inverse Fourier Transform converges to $g$
as a tempered distribution (by continuity of inverse Fourier Transform in $%
T^{\prime }$).

From continuity of the mapping the result follows.

(b) For any $g\in L_{1}$ there exists a sequence of step-functions $g_{n}\in
L_{1}$ such that $\left\Vert g_{n}-g\right\Vert _{L_{1}}\rightarrow 0$
(implying $g_{n}\rightarrow _{T^{\prime }}g);$ for $F$ there is a sequence
of step-functions $F_{n}$ such that $\sup |F_{n}-F|\rightarrow 0$ (implying $%
F_{n}\rightarrow _{T^{\prime }}F).$

Specifically, 
\begin{eqnarray*}
g_{n}(x) &=&\sum_{k=1}^{N}a_{k}I(b_{k}\leq x<b_{k+1})\text{ for }%
b_{1}<...<b_{N}; \\
F_{n}(x) &=&\sum_{j=1}^{N}c_{j}I(d_{j}\leq x)\text{ with }c_{j}>0;\Sigma
c_{j}=1;d_{1}<...<d_{N},
\end{eqnarray*}%
where all the parameters depend on $n.$ The generalized derivative of $F_{n}$
is $f_{n}(x)=\Sigma c_{j}\delta (x-d_{j})$ where $\delta (x-d_{j})$ is a
shifted $\delta -$function: $\int \delta (x-d_{j})\psi (x)dx=\psi (d_{i})$
for $\psi \in T.$ Then $\phi _{n}(\zeta )=\Sigma c_{j}e^{i\zeta d_{j}}.$ The
function $\phi _{n}$ is not integrable (otherwise $f$ would be continuous),
thus $\phi _{n}^{-1}$ satisfies (\ref{cond}). All the parameters depend on $%
n.$

Then 
\begin{eqnarray*}
W_{yn}(v) &=&\sum_{m=1}^{N^{2}}\alpha _{m}I(|v-t_{m}|<\tau _{m}); \\
W_{xyn}(v) &=&\sum_{m=1}^{N^{2}}\alpha _{m}(v-\varepsilon
_{m})I(|v-t_{m}|<\tau _{m}),
\end{eqnarray*}%
where $m$ corresponds to a pair ($k,j)$ and $\alpha
_{m}=a_{k}c_{j};t_{m}=d_{j}+\frac{b_{k}+b_{k+1}}{2},\varepsilon
_{m}=d_{j},\tau _{m}=\frac{b_{k+1}-b_{k}}{2}.$ This represents $W_{yn}$ as a
step-function and $W_{xyn}$ as a piece-wise linear function$.$ The
conditional mean function $W_{y}$ can be consistently estimated in $L_{1}$
by step functions implying existence of a sequence $W_{yn}(v)$ such that $%
\Pr (W_{yn}\rightarrow _{T^{\prime }}W_{y})\rightarrow 1,$ similarly, for
some piece-wise linear $W_{xyn}(v)$ $\Pr (W_{xyn}\rightarrow _{T^{\prime
}}W_{xy})\rightarrow 1$ implying (\ref{converg}), moreover, we can write
(using known Fourier transforms)%
\begin{eqnarray*}
\varepsilon _{yn}(\zeta ) &=&\sum_{k=1}^{N}2\tau _{k}\alpha _{k}\chi
_{k}(\zeta )\mathit{sinc}(\frac{\tau _{k}\zeta }{\pi }); \\
\varepsilon _{xyn}(\zeta ) &=&-i\sum_{k=1}^{N}2\tau _{k}\frac{d}{d\zeta }%
\left[ \alpha _{k}\chi _{k}(\zeta )\mathit{sinc}(\frac{\tau _{k}\zeta }{\pi }%
)\right] -\sum_{k=1}^{N}2\tau _{k}\alpha _{k}\varepsilon _{k}\chi _{k}(\zeta
)\mathit{sinc}(\frac{\tau _{k}\zeta }{\pi }),
\end{eqnarray*}%
where the \textit{sinc}$(x)$ function is defined as $\frac{\sin \pi x}{\pi x}
$ and $\chi _{k}(\zeta )=e^{it_{k}\zeta }.$ The conditions about continuity
and $\varepsilon _{yn}(\zeta )$ non-zero a.e., required in (a) are
satisfied; \textit{(\ref{converg}) }follows from the continuity of the
Fourier transform operator in $T^{\prime }$.$\blacksquare $

Prior to proof of Theorem 2 we make two preliminary observations.

Firstly, under Assumption 5 and 8 (that justifies products of $\gamma _{s}$
and $\overset{\cdot }{\gamma }_{s}$ with $\phi )$ and by Lemma 1 equations (%
\ref{Fy},\ref{Fxy}) in $T^{\prime }$ lead to ($\mathbf{i}^{2}=-1)$:

\begin{eqnarray}
\varepsilon _{y} &=&\varepsilon _{yo}+\varepsilon _{ys}  \label{paraFy} \\
\text{with }\varepsilon _{yo}(\zeta ) &=&\gamma _{o}(\zeta ,\theta ^{\ast
})\phi (\zeta );\varepsilon _{ys}=\gamma _{s}(\theta ^{\ast })\phi (\zeta );
\notag \\
\mathbf{i}\varepsilon _{xy} &=&\mathbf{i}\varepsilon _{xyo}+\mathbf{i}%
\varepsilon _{xyos}+\mathbf{i}\varepsilon _{xys}\text{ }  \label{paraFxy} \\
\text{with }\mathbf{i}\varepsilon _{xyo}(\zeta ) &=&\overset{\cdot }{\gamma }%
_{oo}(\zeta ,\theta ^{\ast })\phi (\zeta ),  \notag \\
\mathbf{i}\varepsilon _{xyos}(\zeta ) &=&\overset{\cdot }{\gamma }%
_{os}(\zeta ,\theta ^{\ast })\phi (\zeta ),\text{ }  \notag \\
\text{and }\mathbf{i}\varepsilon _{xys} &=&\overset{\cdot }{\gamma }%
_{s}(\theta ^{\ast })\phi ,\text{ where }\overset{\cdot }{\gamma }%
_{s}(\theta )\text{ is the generalized derivative of }\gamma _{s}(\theta ). 
\notag
\end{eqnarray}

Second, to construct weighting functions some well-known functions are used.
Denote by $T_{R}\subset T$ the space of real test functions that are Ft of
real-valued functions from $T$; they satisfy $\psi (-\zeta )=\psi (\zeta ).$
A smooth cut-off (or "smudge") function is defined (e.g. in GS or L) as 
\begin{equation*}
f_{cut}(\zeta )=\exp (-\frac{1}{1-\zeta ^{2}})I(\left\vert \zeta \right\vert
<1);
\end{equation*}%
"bump function" is 
\begin{equation*}
f_{bump}(\zeta )=\frac{f_{cut}(\zeta )}{\int_{-1}^{1}f_{cut}(\zeta )d\zeta }.
\end{equation*}%
Consider sets $V,U$ defined as%
\begin{equation}
V=\cup \left( \lbrack a_{i},b_{i}]\cup \left[ -b_{i},-a_{i}\right] \right)
\subset \cup (a_{i}-\varepsilon ,b_{i}+\varepsilon )\cup (-b_{i}-\varepsilon
,-a_{i}+\varepsilon )=U,  \label{uv}
\end{equation}
where $a_{i}\neq b_{i}$ and the intervals and $\varepsilon $ are such that
the only two intervals in $U$ that could intersect would correspond to some $%
i$ with $b_{i}=-a_{i}$; define the function  
\begin{equation*}
f_{V}(\zeta )=I(\left\vert \zeta \right\vert \in V)\ast f_{bump}(\frac{%
2\zeta }{\varepsilon })\frac{2}{\varepsilon }.
\end{equation*}%
This function has the property that it equals $1$ on $V,$ $0$ outside of $U$
and takes values between 0 and 1.

For any $\xi \in R,p\geq 0,\varepsilon >0$ consider a closed set $V_{\xi }=%
\left[ \xi -\alpha ,\xi +\alpha \right] \cup \left[ -\xi -\alpha ,-\xi
+\alpha \right] $  and the function $f_{V_{\xi }}(\zeta ),$ defined above.
Define $f_{\xi ,p}(\zeta )=(\zeta -\xi )^{p}f_{U_{\xi },V_{\xi }}(\zeta ).$
This function has the property that 
\begin{equation*}
\frac{d^{l}f_{\xi ,p,\varepsilon }}{d\zeta ^{l}}(\xi )=(-1)^{l}\frac{%
d^{l}f_{\xi ,p,\varepsilon }}{d\zeta ^{l}}(-\xi )=\left\{ 
\begin{array}{cc}
p! & \text{if }l=p; \\ 
0 & \text{otherwise.}%
\end{array}%
\right. 
\end{equation*}

All the functions, $f_{bump},f_{V},f_{\xi ,p,\varepsilon }$ are in $T_{R}.$

\textbf{Proof of Theorem 2.}

(i) Let $e$ be small enough that closed $e-$neighborhoods of all the points
of \ singularity and discontinuity of $\gamma _{o}$ and $\overset{\cdot }{%
\gamma }_{o}$ do not intersect in $(-\bar{\zeta},\bar{\zeta})$. Define the
union of open intervals that is the compliment to this set in $(-\bar{\zeta},%
\bar{\zeta})$ by $U.$ Construct for a small enough $\varepsilon $ a
corresponding union of closed intervals, $V\subset U$ that can be defined by
(\ref{uv}).  Define $\mu \left( \zeta \right) =f_{V}(\zeta ).$ Then $\frac{%
d^{p}\mu }{d\zeta ^{p}}(s_{l})=0$ for all $s_{l},$ $p;$ integrals $\int 
\overset{\cdot }{\gamma }_{o}(\zeta ,\theta )\mu (\zeta )d\zeta $ and $\int
\gamma _{o}(\zeta ,\theta )\mu (\zeta )d\zeta $ are defined for any $\theta .
$ The inverse Ft's $r_{y}(z,\theta )=Ft^{-1}(\gamma _{o}(-\zeta ,\theta )\mu
(-\zeta ))$ and $r_{xy}(z,\theta )=Ft^{-1}(\mathbf{i}\overset{\cdot }{\gamma 
}_{o}(-\zeta ,\theta )\mu (-\zeta )$ exist. Since $\varepsilon _{yo}(\zeta
)=\gamma _{o}(\zeta ,\theta ^{\ast })\phi (\zeta )$, $\varepsilon
_{xyo}(\zeta )=-\mathbf{i}\overset{\cdot }{\gamma }_{o}(\zeta ,\theta ^{\ast
})\phi (\zeta ),$ $\overset{\cdot }{\gamma }_{o}(\zeta ,\theta )\ $and $%
\varepsilon _{xyo}(\zeta )$ are ordinary locally integrable functions in $%
T^{\prime }$ and $\varepsilon _{yo}(\zeta )$ and $\gamma _{o}(\zeta ,\theta )
$ are continuous and satisfy (\ref{cond}), the products $\varepsilon
_{yo}(\zeta )\overset{\cdot }{\gamma }_{o}(-\zeta ,\theta )$ and $%
\varepsilon _{xyo}(\zeta )\gamma _{o}(-\zeta ,\theta )$ are well defined in $%
T^{\prime }$ $.$ Thus the integral (where $\mu \in T_{R})$ 
\begin{equation*}
\int \left[ \varepsilon _{yo}(\zeta )\mathbf{i}\overset{\cdot }{\gamma }%
_{o}(-\zeta ,\theta )\mu (-\zeta )+\varepsilon _{xyo}(\zeta )\gamma
_{o}(-\zeta ,\theta )\mu (-\zeta )\right] d\zeta 
\end{equation*}%
exists. Since $\varepsilon _{yo}(\zeta )=\gamma _{o}(\zeta ,\theta ^{\ast
})\phi (\zeta )$, $\varepsilon _{xyo}(\zeta )=-\mathbf{i}\overset{\cdot }{%
\gamma }_{o}(\zeta ,\theta ^{\ast })\phi (\zeta )$ the value of the integral
is zero for $\theta =\theta ^{\ast }.$ Moreover, because the functions $\mu $
are zero together with all the derivatives at singularity points, $%
\varepsilon _{yo}$ can be replaced by $\varepsilon _{y}$  providing:%
\begin{eqnarray*}
&&\int \left[ \varepsilon _{yo}(\zeta )\mathbf{i}\overset{\cdot }{\gamma }%
_{o}(-\zeta ,\theta )\mu (-\zeta )+\varepsilon _{xyo}(\zeta )\gamma
_{o}(-\zeta ,\theta )\mu (-\zeta )\right] d\zeta  \\
&=&(\mathbf{i}\varepsilon _{y}\overset{\cdot }{\gamma }_{o}(-\zeta ,\theta
),\mu (-\zeta ))+(\varepsilon _{xy}\gamma _{o}(-\zeta ,\theta ),\mu (-\zeta
))
\end{eqnarray*}%
By applying Parseval identity to generalized functions this leads to  
\begin{eqnarray*}
&&(W_{y},r_{xy}(\theta ))+(W_{xy},r_{y}(\theta )) \\
&=&\int \left[ W_{y}(z)r_{xy}(z,\theta )+W_{xy}(z)r_{y}(z,\theta )\right] dz
\end{eqnarray*}%
where the functionals are expressed via integrals for ordinary locally
integrable functions.

Multiplying and dividing by the non-zero function $p(z)$ does not change the
integral. Then by law of iterated expectations 
\begin{equation*}
\int \frac{1}{p(z)}E_{|z}\left( Yr_{xy}(z,\theta )+XYr_{y}(z,\theta )\right)
p(z)dz=E\left( \frac{Yr_{xy}(z,\theta )+XYr_{y}(z,\theta )}{p(z)}\right) .
\end{equation*}

This concludes the proof of (i).

(ii) By Assumption 5(iii) there exists a sequence $\xi _{n}\rightarrow 0$
such that $\gamma _{o}(\zeta ,\theta )\neq 0$ for $\zeta :\left\vert \zeta
-\xi _{n}\right\vert <\varepsilon _{n}<\left\vert \xi _{n}\right\vert $ and
is continuous in those intervals; without loss of generality assume that $%
\zeta \in U$ defined in (i). Consider the function 
\begin{equation*}
\mu _{n}(\zeta )=\frac{1}{2}\{f_{bump}(\frac{\zeta -\xi _{n}}{\varepsilon
_{n}})+f_{bump}(\frac{\zeta +\xi _{n}}{\varepsilon _{n}})\}
\end{equation*}%
The function $\frac{\mu _{n}(-\zeta )}{\overline{\gamma _{o}(-\zeta ,\theta )%
}}$ is a continuous function with bounded support. Set $r_{1yn}(z,\theta
)=Ft^{-1}(\frac{\mu _{n}(-\zeta )}{\overline{\gamma _{o}(-\zeta ,\theta )}}%
). $Then for any $n$ we get

\begin{eqnarray*}
\int \varepsilon _{yo}(\zeta )\overline{\gamma _{o}(-\zeta ,\theta )}%
^{-1}\mu _{n}(-\zeta )d\zeta  &=&(\varepsilon _{y}\cdot \overline{\gamma
_{o}(-\zeta ,\theta )}^{-1},\mu _{n}(-\zeta )) \\
&=&\int E(Y|z)Ft^{-1}(\overline{\gamma _{o}(-\zeta ,\theta })^{-1}\mu
_{n}(-\zeta ))dz \\
&=&E\left( \frac{Yr_{y1n}(z,\theta )}{p(z)}\right) ,
\end{eqnarray*}%
where the first equality follows from the fact that ($\varepsilon _{ys}%
\overline{\gamma _{o}(-\zeta ,\theta )}^{-1},\mu _{n}(-\zeta ))=0$ (since $%
\zeta \in U),$ the second by Parseval identity and the third by multiplying
and dividing by $p(z)>0$ and iterated expectation; the integral exists for
each $n$. For $\theta ^{\ast }$ we get ($\gamma _{o}(\zeta )=\overline{%
\gamma _{o}(-\zeta )})$ 
\begin{eqnarray*}
E\left( \frac{Yr_{y1n}(z,\theta ^{\ast })}{p(z)}\right)  &=&\int \varepsilon
_{yo}(\zeta )\gamma _{o}(\zeta ,\theta ^{\ast })^{-1}\mu _{n}(-\zeta )d\zeta
=\int \phi (\zeta )\mu _{n}(-\zeta )d\zeta  \\
&=&\frac{1}{2}\left[ \phi (\xi _{n})+\phi (-\xi _{n})\right] +O(\varepsilon
_{n})
\end{eqnarray*}%
This converges to $\phi (0)=1.$

(iii) Consider any $s_{l}\geq 0$. Below all relevant functions are
subscripted by $l$. 

For $\varepsilon $ as defined in (i) define the function $\mu _{l,i}(\zeta
)=f_{s_{l},i,\varepsilon }(\zeta )\in T_{R}$, then $\mu _{l,i}^{(i)}(0)\neq
0,$ but $\mu _{l,i}^{(k)}(0)=0,$ $k=0,...i-1,i+1,...\bar{k}+1$ and support
of $\mu _{l,i}$ is given by $I(\left\vert \zeta -s_{l}\right\vert
<\varepsilon )+I(\left\vert \zeta +s_{l}\right\vert <\varepsilon );$ denote
the derivative of $\mu _{l,i}$ by $\mu _{l,i}^{\prime }$. For a sequence $%
\varepsilon _{n}\rightarrow 0$ consider $f_{V_{n}}(\zeta )$ for $%
U_{n}=\{\zeta :|\zeta -s_{l}|<\varepsilon _{n}\}\cup \{\zeta :\left\vert
\zeta +s_{l}\right\vert <\varepsilon _{n}\};V_{n}=\{\zeta :|\zeta
-s_{l}|\leq \frac{\varepsilon _{n}}{2}\}\cup \{\zeta :\left\vert \zeta
+s_{l}\right\vert <\frac{\varepsilon _{n}}{2}\}$ and define $\mu
_{l,i,n}(\zeta )=\mu _{li}(\zeta )f_{U_{n},V_{n}}(\zeta )).$ The functions $%
\mu _{\cdot }$ are in $T_{R}.$ Denote by $r_{xys,l,i,n}(z)$ the inverse Ft: $%
Ft^{-1}(\mu _{l,i,n}(-\zeta ))$ and by $r_{ys,l,i,n}(z)$ the inverse Ft: $%
Ft^{-1}(\mathbf{i}\mu _{l,i,n}^{\prime }(-\zeta ));$ they exist in $T.$ The
vector $r_{xys,l,n}(z)$ is defined to have $r_{xys,l,i,n}(z)$ as its $i-$th
component; vector $r_{ys,l,n}(z)$ is defined similarly. Define by $M_{l}$
the diagonal matrix with non-zero diagonal entries $\left\{ M_{l}\right\}
_{ii}=\mu _{l,i,n}^{(i)}(0)\equiv \mu _{l,i}^{(i)}(0)$, $i=0,...\bar{k}.$

Consider now the vector $(\varepsilon _{y},\mu _{l,n}^{\prime })$ with
components $(\varepsilon _{y},\mu _{l,i,n}^{\prime }(-\zeta ))$ and $%
(\varepsilon _{xy},\mu _{l,n})$ with $(\varepsilon _{xy},\mu _{l,i,n}(-\zeta
))$. Since the matrices $\Gamma _{y}(s_{l},\theta )$, $\Gamma
_{xy}(s_{l},\theta )$ and $M_{l}$ are invertible the expression 
\begin{equation}
\Gamma _{y}(s_{l},\theta )^{-1}M_{l}^{-1}(\varepsilon _{y},\mu
_{l,n}^{\prime })+\Gamma _{xy}(s_{l},\theta )^{-1}M_{l}^{-1}(\varepsilon
_{xy},\mu _{l,n})  \label{smom}
\end{equation}%
is finite for every $n$. By Parseval identity 
\begin{eqnarray*}
(\varepsilon _{y}(\zeta ),\mu _{l,i,n}^{\prime }(-\zeta ))
&=&(W_{y}(z),r_{y,l,i,n}(z)) \\
&=&\int W_{y}(z)r_{y,l,i,n}(z)dz,
\end{eqnarray*}%
the last equality follows since $W_{y}$ is locally integrable. Thus by
arguments similar to those in (i) and (ii) this integral is $E\frac{%
Yr_{y,l,i,n}(z)}{p(z)}$ so that $(\varepsilon _{y},\mu _{l,n}^{\prime })=E%
\frac{Yr_{y,l,n}(z)}{p(z)}$ and analogously $(\varepsilon _{xy},\mu _{l,n})=E%
\frac{XYr_{xy,l,n}(z)}{p(z)}.$ We need to establish that limits as $%
n\rightarrow \infty $ exist. First, note that 
\begin{eqnarray*}
\left\vert \int \varepsilon _{yo}(\zeta )\mu _{l,i,n}^{\prime }(-\zeta
)d\zeta \right\vert  &=&\left\vert \int \varepsilon _{yo}(\zeta )f_{\zeta
_{l},i,\varepsilon }(-\zeta )f_{U_{n},V_{n}}(-\zeta )d\zeta \right\vert  \\
&\leq &\underset{U}{\max }\left\vert \varepsilon _{yo}(\zeta )f_{\zeta
_{l},i,\varepsilon }(-\zeta )\right\vert 2\varepsilon _{n}
\end{eqnarray*}%
and goes to zero; 
\begin{equation}
(\varepsilon _{ys},\mu _{l,i,n}^{\prime }(-\zeta ))=\Sigma _{k\geq i}^{\bar{k%
}}\gamma (s_{l},\theta ^{\ast })(-1)^{i}\left( 
\begin{array}{c}
k+i-1 \\ 
i-1%
\end{array}%
\right) \mu _{l,i}^{(i)}(s_{l})\phi ^{(k-i+1)}(s_{l})  \label{sexp}
\end{equation}%
and does not depend on $n,$ finally, $\varepsilon _{y}=\varepsilon
_{yo}+\varepsilon _{ys},$ so the limit exists. By a similar representation
for $-(\varepsilon _{xys},\mu _{l,i,n}(-\zeta ))$ existence of (\ref%
{gammasshape}) is established. For $\theta =\theta ^{\ast }$ using (\ref%
{paraFy},\ref{paraFxy}) for $\varepsilon _{ys}\ $and $\varepsilon _{xys}$ in
($\ref{smom})$ leads to

\begin{equation*}
\Gamma _{y}(s_{l},\theta ^{\ast })^{-1}M_{l}^{-1}(\varepsilon _{ys},\mu
_{l,n}^{\prime })+\Gamma _{xy}(s_{l},\theta ^{\ast
})^{-1}M_{l}^{-1}(\varepsilon _{xys},\mu _{l,n})=0.
\end{equation*}%
Note that the same considerations apply to singularity at $-s_{l}$ with the
difference that the $\Gamma _{.}(-s_{l},\theta )$ matrices now are complex
conjugate to $\Gamma _{.}(s_{l},\theta ).$ Combining provides the real part
in (\ref{gammasshape}).

(iv) Consider the first component of $\Gamma _{y}(s_{l},\theta
)^{-1}M_{l}^{-1}(\varepsilon _{ys},\mu _{l,n}^{\prime })$ with $\mu
_{l,n}^{\prime },M_{l}$ defined in (iii); this component is $\{\Gamma
_{y}(s_{l},\theta )^{-1}M_{l}^{-1}\}_{11}(\varepsilon _{ys},\mu
_{l,1,n}^{\prime }).$ Note that $\mu _{l,1,n}^{\prime }=\mu _{l,0,n},$
recall that $\mu \in T_{R}.$ We see that $\phi (s_{l})$ equals $\lim
\{\Gamma _{y}(s_{l},\theta ^{\ast })^{-1}M_{l}^{-1}\}_{11}\int \varepsilon
_{ys}\mu _{l,0,n}d\zeta .$ Define 
\begin{equation*}
r_{ysl,1,n}(z,\theta )=\{\Gamma _{y}(s_{l},\theta
)^{-1}M_{l}^{-1}\}_{11}Ft^{-1}(\mu _{l,0,n}(-\zeta )).
\end{equation*}%
Similarly to above by Parseval identity $\{\Gamma _{y}(s_{l},\theta
)^{-1}M_{l}^{-1}\}_{11}(\varepsilon _{ys},\mu _{l,0,n})=E(\frac{%
Yr_{ysl,1,n}(z,\theta )}{p(z)})$ and $\phi (s_{l})=\lim E(\frac{%
Yr_{ysl,1,n}(z,\theta )}{p(z)}).$ For $s_{0}=0$ we have $\phi (0)=1$. Thus (%
\ref{gamma0scale}) follows.

Consider now for $s_{l}\neq 0$ the function 
\begin{equation*}
\mu _{l,n}(\zeta )=\frac{1}{2}\{f_{bump}(\frac{\zeta -s_{l}-\xi _{n}}{%
\varepsilon _{n}})+f_{bump}(\frac{\zeta -s_{l}+\xi _{n}}{\varepsilon _{n}})\}
\end{equation*}%
similar to the one in (ii) and define $r_{slo,n}(z)=Ft^{-1}(\frac{\mu
_{l,n}(-\zeta )}{\overline{\gamma _{o}(-\zeta ,\theta )}}).$ For this
function $E(\frac{Yr_{slo,n}(z,\theta )}{p(z)})=\int \varepsilon _{y}(\zeta )%
\frac{\mu _{l,n}(-\zeta )}{\overline{\gamma _{o}(-\zeta ,\theta )}}d\zeta $
exists and at $\theta ^{\ast }$ converges to $\phi (s_{l})$. Thus (\ref%
{gammasscale}) follows.$\blacksquare $

Proof of Theorem 3.

Let the vector $Q(z,\theta )$ denote the vector of functions for which
expectations are taken in $E(Q);$ partition $Q(z,\theta )$ into $%
Q_{I}(z,\theta )$ corresponding to expressions in (\ref{gammaoshape}, \ref%
{gammaoscale}) and $Q_{II}(z,\theta )$ for (\ref{gammasshape}, \ref%
{gammasscale}). Then the matrix $\frac{\partial }{\partial \theta ^{T}}%
E(Q(z,\theta ))$ is a block matrix 
\begin{equation*}
\left( 
\begin{array}{cc}
\frac{\partial }{\partial \theta _{I}^{T}}E(Q_{I}(z,\theta )) & \cdot \\ 
\cdot & \frac{\partial }{\partial \theta _{II}^{T}}E(Q_{II}(z,\theta ))%
\end{array}%
\right)
\end{equation*}%
and it is sufficient to show that $\frac{\partial }{\partial \theta _{I}^{T}}%
E(Q_{I}(z,\theta ))$ has rank $m_{I}$ and $\frac{\partial }{\partial \theta
_{II}^{T}}E(Q_{II}(z,\theta ))$ has rank $m_{II}.$

For $\theta _{I}$ first note that interchange of differentiation with
respect to the parameter and integration (taking expected value) for $\frac{%
\partial }{\partial \theta _{I}^{T}}EQ_{I}(z,\theta )$ follows from
continuity in $\zeta $ of all the functions in the integrals and their
continuous differentiability with respect to $\theta ,$ so that $\frac{%
\partial }{\partial \theta _{I}}E(Q_{I}(z,\theta ))=E(\frac{\partial }{%
\partial \theta _{I}}Q_{I}(z,\theta )).$ One can choose $m_{I}$ functions $%
\mu $ defined in proofs of Theorem 2(i,ii) that are functionally independent
and under Assumption 6 the corresponding $m_{I}$ conditions of type $E(\frac{%
\partial }{\partial \theta _{I}}Q_{I}(z,\theta ))$ will provide a rank $m_{I}
$ submatrix.

If for the functions $\mu $ in expressions $Q_{II}(z,\theta )$ in (iii,iv)
of Proof of Theorem 2 the matrix $\frac{\partial }{\partial \theta _{II}}%
E(Q_{s}(z,\theta ))$ has rank less than $m_{II}$ consider varying the
functions $\mu _{\cdot }$ for all possible values of non-zero derivatives at
the points $s_{l};$ the rank cannot be deficient over all such choices
without violation of Assumption 6.$\blacksquare $

{\LARGE Appendix B}

Three main problems with (S) are listed below.

\textbf{1. The issue of decomposition. }

(S) claims that any generalized function in $T^{\prime }$ can be decomposed
into a sum of an ordinary function and a singular function; this
decomposition is used in formula (13) of S., Theorem 1.

There is no proof of existence of such a decomposition in the literature, as
the example of the function in Section 2.1 in (\ref{E2}) shows no unique
decomposition into a singular and regular ordinary function exists in $%
T^{\prime }$ (nor in $D^{\prime })$. The attempted proof in (S,
Supplementary material, p.3) is incorrect.

Indeed it states: "The result directly follows from the fact that every
generalized function can be written as the derivative of order $k\in N$ of
some continuous function $c(t)$ (Theorem III in Temple (1963) establishes
this for a class of generalized functions including those considered here as
a particular case). \textit{[-my comment: this is correct]} At every point $%
t $ where $c(t)$ is $k$ times differentiable in the usual sense, the
generalized function can be written as an ordinary function, while at every
point where $c(t)$ is not $k$ times differentiable, a delta function
derivative is created in the differentiation process.\textit{[my comment:
this is incorrect, see (a) below]} The fact that the two pieces are
additively separable follows from the linear nature of the space of
generalized functions.\textit{\ [my comment: this is incorrectly applied:
see (b) below]" }

(a). Consider the function $b(\zeta )=\left\vert \zeta \right\vert ^{-\frac{3%
}{2}};$ it is a "weak" (or generalized) second derivative of the continuous
function $c(\zeta )=-4\left\vert \zeta \right\vert ^{\frac{1}{2}}$. The
first weak derivative, $-2\left\vert \zeta \right\vert ^{-\frac{1}{2}%
}sign(\zeta ),$ is an ordinary function that is summable and so gives a
regular functional and is an ordinary function that is at the same time a
generalized function. At point $\zeta =0$ it is not differentiable in the
ordinary sense; yet \textit{no delta-function or its derivative appears}.

(b) Any generalized function is either regular or singular; the "ordinary"
function $b(\zeta )$ above is a singular generalized function (see (\ref{E2}%
)). Since generalized functions are generally not defined pointwise a
pointwise argument cannot be helpful.

An additional assumption would have to be used to establish formula (13) in
Theorem 1 of (S).

\textbf{2. Validity of products.}

Validity of products in the space of generalized functions needs to be
established to provide a correct proof of the identification result. Neither
the paper nor the supplementary material in (S) provides a complete correct
proof indicating in which space of generalized functions the multiplication
operations are valid; in fact as Lemma 1 here shows multiplication may not
be valid in $T^{\prime }$ under the Assumptions (despite the claim in (S)). 

\textbf{3. Definition 2 in (S) leads to inappropriate weights.}

Def. 2 proposes the function%
\begin{equation*}
\mu (\zeta )=\dsum\limits_{k=0}^{\infty }\frac{1}{k!}d^{k}(\zeta ^{k}\lambda
(\zeta ))/d\zeta ^{k}
\end{equation*}%
for $\lambda $ that satisfies S, Def. 1 ($\lambda $ is an analytic
function). We have

\begin{eqnarray*}
&&\dsum\limits_{k=0}^{\infty }\frac{1}{k!}d^{k}(\zeta ^{k}\lambda (\zeta
))/d\zeta ^{k} \\
&=&\dsum\limits_{k=0}^{\infty }\frac{1}{k!}\dsum\limits_{i=0}^{k}C_{i}^{k}%
\left( d^{i}\zeta ^{k}/d\zeta ^{i}\right) \left( d^{k-i}\lambda (\zeta
)/d\zeta ^{k-i}\right) \\
&=&\dsum\limits_{k=0}^{\infty }\frac{1}{k!}\dsum\limits_{i=0}^{k}C_{i}^{k}%
\frac{k!}{(k-i)!}\zeta ^{k-i}\left( d^{k-i}\lambda (\zeta )/d\zeta
^{k-i}\right) \\
&=&\dsum\limits_{k=0}^{\infty }\dsum\limits_{i=0}^{k}C_{i}^{k}\frac{1}{i!}%
\zeta ^{i}\left( d^{i}\lambda (\zeta )/d\zeta ^{i}\right) .
\end{eqnarray*}

Consider the values and derivatives of the function $\mu $ at zero; they are
zero and thus are not suitable for the weighting functions. Indeed 
\begin{eqnarray*}
\mu (0) &=&\dsum\limits_{k=0}^{\infty }\lambda (0),\text{ so }\lambda (0)=0%
\text{ and }\mu (0)=0 \\
\mu ^{\prime }(0) &=&\dsum\limits_{k=0}^{\infty
}\dsum\limits_{i=0}^{k}C_{i}^{k}\frac{1}{i!}[\zeta ^{i-1}\left( d^{i}\lambda
(\zeta )/d\zeta ^{i}\right) +\zeta ^{i}\left( d^{i+1}\lambda (\zeta )/d\zeta
^{i+1}\right) |_{\zeta =0} \\
&=&\dsum\limits_{k=0}^{\infty }[\lambda ^{\prime }(0)+k\lambda ^{\prime
}(0)],\text{ so }\lambda ^{\prime }(0)=0,\text{ and }\mu ^{\prime }(0)=0, \\
\text{etc., implying }\mu ^{(k)}(0) &=&0\text{ for any }k.
\end{eqnarray*}

\bigskip

\bigskip

\bigskip \pagebreak

\end{document}